\documentclass[showkeys,showpacs]{revtex4}
\usepackage{graphicx}
\usepackage{amsmath}
\usepackage{amssymb}
\usepackage{color}
\usepackage{enumerate}

\begin{document}
\title{Topology change in quantum gravity and Ricci flows}

\author{Vladimir Dzhunushaliev}
\email{vdzhunus@krsu.edu.kg}
\affiliation{Department of Physics and Microelectronic
Engineering, Kyrgyz-Russian Slavic University, Bishkek, Kievskaya Str.
44, 720021, Kyrgyz Republic \\ 
Institute of Physicotechnical Problems and Material Science of the NAS
of the Kyrgyz Republic, 265 a, Chui Street, Bishkek, 720071,  Kyrgyz Republic \\
}

\author{Nurzhan Serikbayev}
\affiliation{Dept. Gen. Theor. Phys., Eurasian National University, Astana, 010008, Kazakhstan}

\author{Ratbay Myrzakulov}
\affiliation{Dept. Gen. Theor. Phys., Eurasian National University, Astana, 010008, Kazakhstan
}
\email{cnlpmyra1954@yahoo.com}

\begin{abstract}
The topology change in quantum gravity is modeled by a Ricci flow. In this approach we offer to consider the Ricci flow as a statistical system. The metric in the Ricci flow enumerated by a parameter $\lambda$ is a microscopical statistical state. The probability of every microscopical state is determined by the parameter $\lambda$. Numerically the Ricci flow starting from a static wormhole filled with a phantom Sine-Gordon scalar field is investigated. 
\end{abstract}

\date{\today}

\maketitle

\section{Introduction}

\begin{figure}[h]
\fbox{
	\includegraphics[height=5cm,width=5cm]{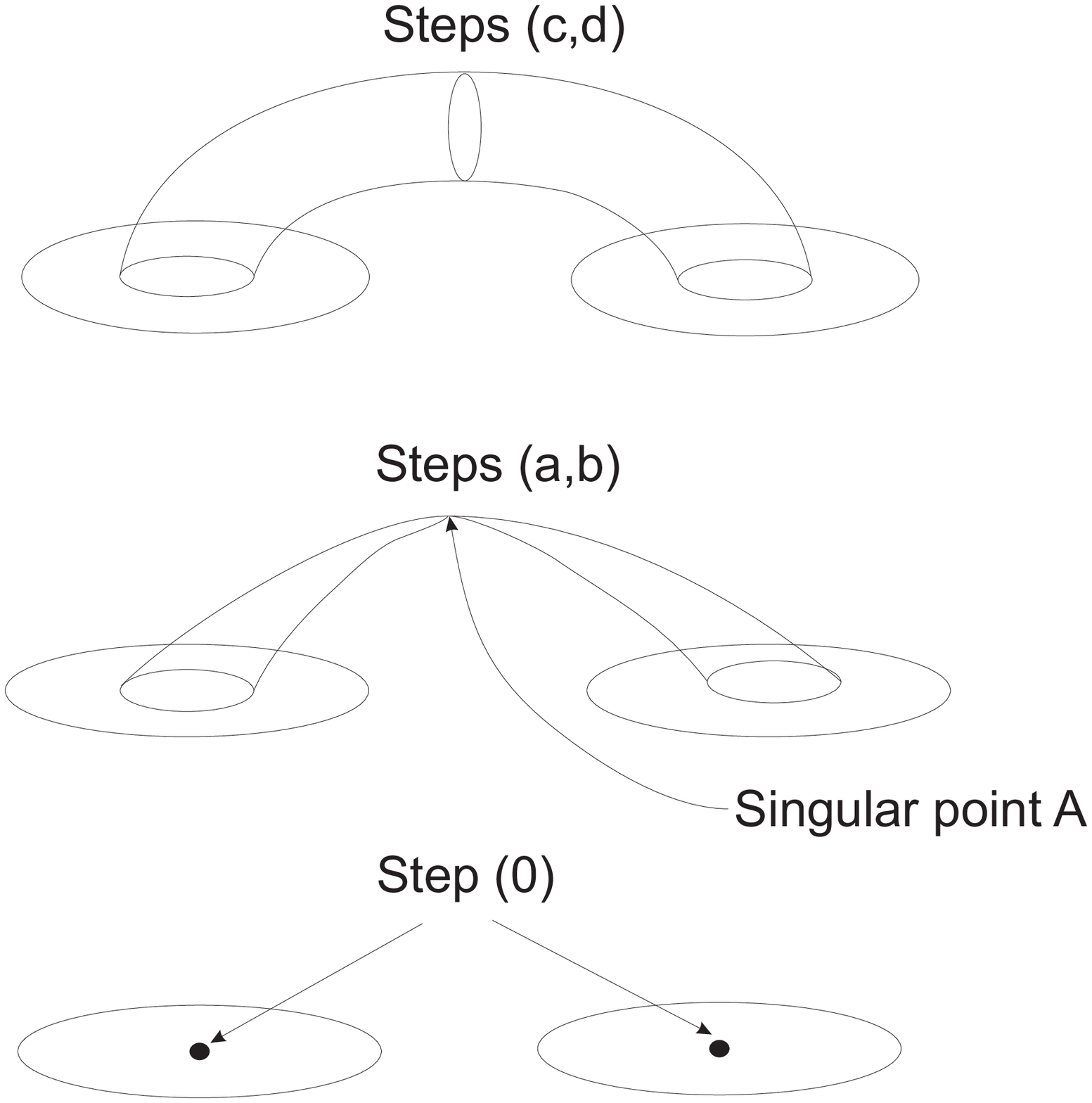}}
		\caption{Step(0) is two initial points. Steps (a,b) are the identification of two points with the subsequent deletion of the obtained point. Steps (c,d) are a boundary is pasted in with the subsequent blow up. }
\label{mors}
\end{figure}

One of the most interesting questions in a hypothetical quantum gravity is the question about topology change. The essence of the problem is that on a microscopical level (where the typical length of a phenomenon lies in the region of Planck length $\approx 10^{-33}$cm) may take place the change of a space topology. This process is called as a spacetime foam \cite{wheel1}. Wheeler has proposed the idea that on the Planck level the metric fluctuations are so big that the change of space topology happens. This process occurs in a few steps: (a) two points identified; (b) the obtained point is deleted; (c) at the boundary of obtained space a sphere is pasted in; (d) the sphere is blow up (see Fig. \ref{mors}). 

The main problem in this model is that in during of the topology change a singularity may be created \cite{geroch0}. The origin of the singularity is that in during of the process some points are deleted and some points are inserted. It follows to the singularity in the same way as the creation of universe from the Big Bang bring to the cosmological singularity. There have been a
number of investigations aiming to demonstrate the feasibility of topology change \cite{borde}, \cite{horowitz}, \cite{dowkergarcia}, \cite{Sorkin}. In Ref. \cite{Csizmadia:2009dm} the numerical investigation of topology change in Einstein - matter system is made. Nevertheless, the investigations in this direction are still far from the end. We would like to emphasise that some of our findings provide a qualitative (and in the future quantitative) support to the  speculations concerning the existence of topology changes. 

In the differential topology such process is investigated in order to understand the topological structure of a manifold. The process is known as a topological surgery \cite{Hirsch}. In recent years a great success was achieved in this direction. The modern mathematical tools for the investigation of the topology surgery is Ricci flows. The Ricci flows were introduced by Hamilton \cite{Hamilton} over 25 years ago. They play an important role in the proof of the Poincare conjecture \cite{Perelman}. The Ricci flows are differential equations for the metric on a manifold $\mathcal M$. The equations describe the creation on the manifold a singularity (or singularities) for a finite parameter $\lambda_0$. Fig.~\ref{mors} (Step (a,b)) gives a schematic picture of the partially singular metric $g(\lambda_0)$ on the manifold $\mathcal M$. The Ricci flow evolves from Step (c,d) to Step (a,b) in Fig. \ref{mors}. The metric $g(\lambda_0)$ is smooth almost everywhere excluding the point where two initial points were identified. Strictly speaking the curvature is locally bounded  for almost all $\lambda$ but blows-up as 
$\lambda \rightarrow \lambda_0$ at one point or more than one points. 
\par 
Let the metric $g(\lambda)$ is smooth on a domain $\Omega \subset \mathcal M$. But for 
$\lambda \rightarrow \lambda_0$ the metric becomes singular at one or more that one points (point $A$ in Fig. \ref{mors}). If $\Omega \neq \emptyset$, then the main point is that small neighborhoods of the boundary $\partial \Omega$ consist of horns. A horn is a metric on $S^2 \times [0, \delta]$ where the $S^2$ factor is approximately round of radius $\rho(r)$, with $\rho(r)$ small and $\rho(r)/r \rightarrow 0$ as $r \rightarrow 0$. Fig.~\ref{mors} represents a partially singular metric on the smooth manifold $S^2 \times I$ consisting of a pair horns joined by a degenerate metric: $g(\lambda = \lambda_0)\rvert_{x=A}  = \infty$. 
\par
The main idea of this work is to model the topology change in a space using the Ricci flow mathematics. We offer the idea that for every $\lambda$ the 3D space-like metric $g(\lambda)$ is realised with some probability $\rho(\lambda)$ where the parameter $\lambda$ describes the evolution of the metric $g$ under the Ricci flow. Usually such probability is calculated from the path integral. The idea presented here is that the probability is connected with a Perelman's functional $\mathcal W$ on a \textcolor{blue}{\emph{rescaled}} Ricci flow as $\rho = f \left( \frac{d \mathcal W}{d \lambda} \right)$ in the consequence of the property 
$\frac{d \mathcal W}{d \lambda} \geq 0$. Then the Ricci flow is a \textcolor{blue}{\emph{statistical}} system where every metric $g(\lambda)$ is a \textcolor{blue}{\emph{microscopical}} state. 

In this work we continue to investigate the idea that the topology change in quantum gravity is connected with the Ricci flows \cite{Dzhunushaliev:2008cz}. Here we consider another type of an initial wormhole from which the Ricci flow is started. The initial wormhole is not asymptotically flat and has anti de-Sitter asymptotic. The physical idea offered in Ref. \cite{Dzhunushaliev:2008cz} is following. Let us we have a static wormhole solution in general relativity. Then the Ricci flow starting from the initial wormhole describes the topology change. The Ricci flow is the sequence of 3D metrics enumerated by the parameter $\lambda$. The main idea is that \textcolor{blue}{\emph{the Ricci flow is a statistical system and every state with the metric $g(\lambda)$ is a microscopical state with the probability determined by the functional \eqref{ric-70}.}}

\section{Short introduction to Ricci flows}
\label{ricci}

Now we would like to present a short introduction to Ricci flows following to Ref.~\cite{topping}. Ricci flow is a means of processing the metric $g_{ab}$ by allowing it to evolve under 
\begin{equation}
	\frac{\partial g_{ab}(x^c, \lambda)}{\partial \lambda} = -2 R_{ab}(x^c, \lambda)
\label{1-10}
\end{equation}
where $R_{ab}$ is the Ricci curvature; $\lambda$ is a parameter; 
$a,b,c = 1,2, , \cdots , n = \dim \mathcal M$; $x^a$ are the local coordinates on a manifold $\mathcal M$. The Ricci flow describes the evolution of the metric $g_{ab}$ in during of the parameter $\lambda$. 
\par 
Let us introduce the Perelman's functional
\begin{equation}
	\mathcal W(g_{ab}, f, \tau) = 
	\int \left[
		\tau \left( R + \left| f \right|^2 \right) + f - n
	\right] \, u \; dV
\label{ric-20}
\end{equation}
where $f: \mathcal M \rightarrow \mathbb R$ is a smooth function; $R$ is the Ricci scalar; $\tau > 0$ is a scale parameter; $n = \mathrm{dim} \mathcal M$, and $u$ is defined by 
\begin{equation}
	u= (4 \pi \tau)^{-n/2} e^{-f}. 
\label{ric-30}
\end{equation}
One can show that if $\mathcal M$ is closed, and $g_{ab}, f$ and $\tau$ satisfy following equations 
\begin{eqnarray}
	\frac{\partial g_{ab}}{\partial \lambda} &=& -2 R_{ab},
\label{ric-40}\\
	\frac{d \tau}{d \lambda} &=& -1,
\label{ric-50}\\
	\frac{\partial f}{\partial \lambda} &=& 
	- \Delta f + \left| \nabla f \right|^2 - R + 
	\frac{n}{2 \tau}
\label{ric-60}
\end{eqnarray}
then the functional $\mathcal W$ increases according to 
\begin{equation}
	\frac{d}{d \lambda} \mathcal W(g_{ab}, f, \tau) = 
	2 \tau \int \left|
		R_{ab} + \frac{\partial^2 f}{\partial x^a \partial x^b} - 
		\frac{g_{ab}}{2 \tau}
	\right|^2 u dV \geq 0.
\label{ric-70}
\end{equation}
Under evolution \eqref{ric-40}-\eqref{ric-60} 
\begin{equation}
	\frac{d}{d \lambda} \int u dV = 0.
\label{ric-80}
\end{equation}
It means that 
\begin{equation}
	\int u dV = const
\label{ric-90}
\end{equation}
and $u(\lambda)$ represents the probability density of a particle evolving under Brownian motion, backwards in time. It allows us to define the classical, or ``Boltzman - Shannon'' entropy 
\begin{equation}
	S = - \int \limits_{\mathcal M} u \; \mathrm{ln} u \; dV 
\label{ric-100}
\end{equation}
or a renormalized version of the classical entropy 
\begin{equation}
	\tilde{S} = S - \frac{n}{2}  \left \{
		1 + \mathrm{ln} \left[ 4 \pi (\lambda_0 - \lambda) \right] 
	\right \}.
\label{ric-110}
\end{equation}
The $\mathcal W-$functional applied to this backwards Brownian diffusion on a Ricci flow also arises via the renormalized classical entropy $\tilde S$ 
\begin{equation}
	\mathcal W(\lambda) = - \frac{d}{d \lambda} \left( \tau \tilde S \right). 
\label{ric-120}
\end{equation}

\section{Topology change with Ricci flow}

In this section we would like to show that starting from an initial wormhole solution in general relativity one can obtain the sequence of 3D metric. The sequence converges to the space with a point where the metric has a singularity in the spirit of Ricci flows. Such sequence can be illuminating by Steps~(c,d)~$\rightarrow$~Steps~(a,b) in Fig.~\ref{mors}. In the next subsection we will present the wormhole solution filled with a phantom scalar field. 

\subsection{Wormhole solution with a phantom scalar field}
\label{wh}

\begin{figure}[ht]
\begin{center}
	\fbox{
  \includegraphics[width=.5\linewidth]{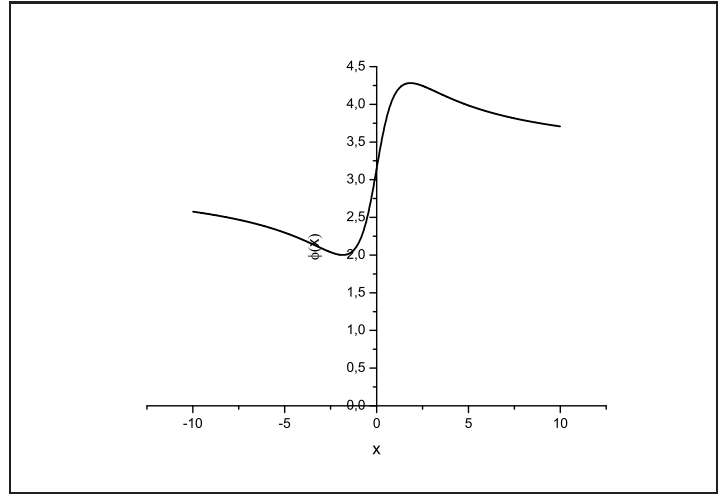}
  }
 	\caption{The profile of the scalar field $\phi(x)$. $\beta=1$. Asymptotically 
 	$\phi\rightarrow \pi$.}
\label{phi_wh_sine}
\end{center}
\end{figure}

\begin{figure}[h]
\begin{minipage}[t]{.45\linewidth}
  \begin{center}
  \fbox{
  \includegraphics[height=1.0\linewidth,width=1.0\linewidth]{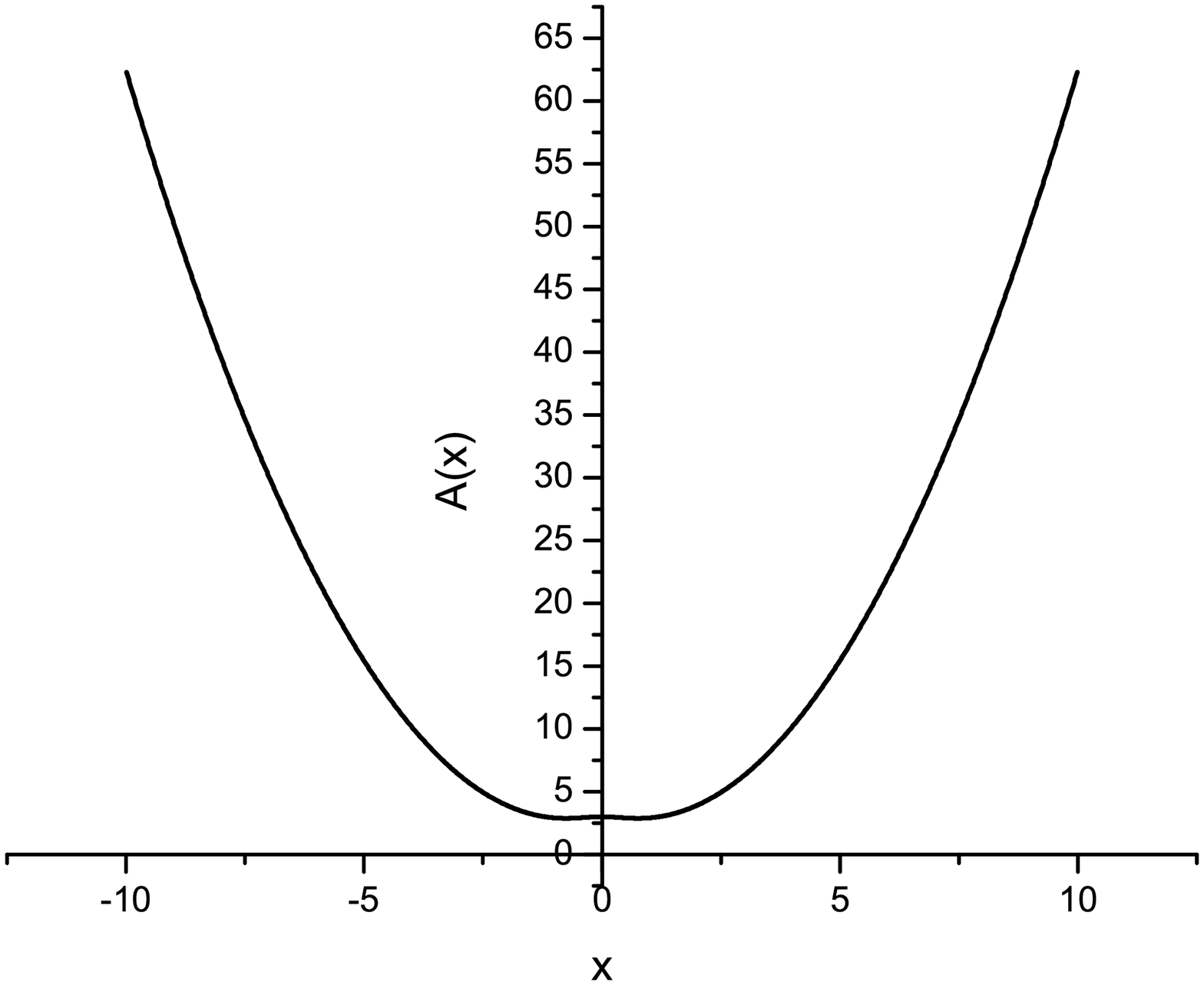}
  }
 	\caption{The profile of $A(x)$.}
	\label{A_wh_sine}
  \end{center}
\end{minipage}\hfill
\begin{minipage}[t]{.45\linewidth}
  \begin{center}
  \fbox{
  \includegraphics[height=1.0\linewidth,width=1.0\linewidth]{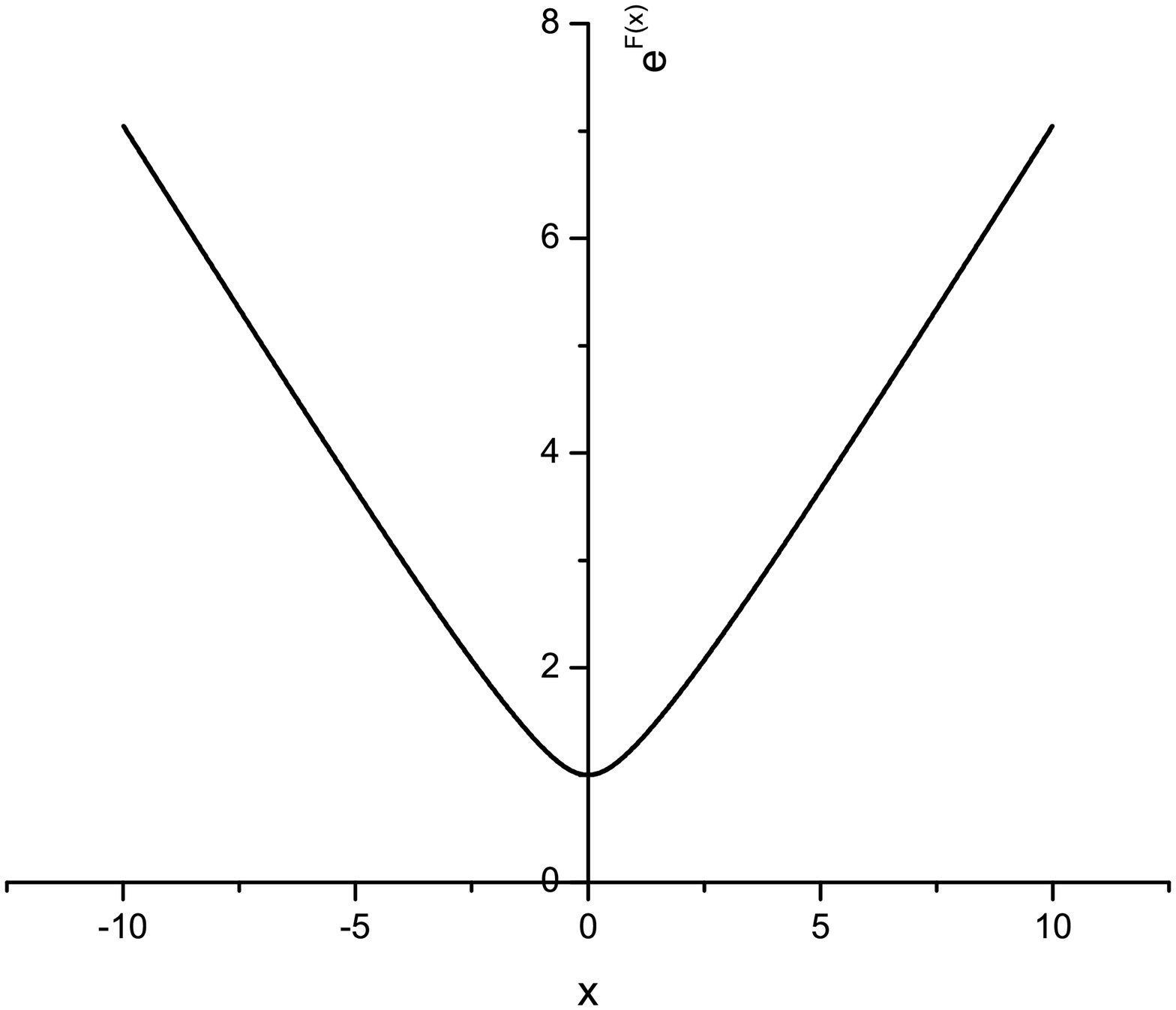}
  }
 	\caption{The profile of $e^{F(x)}$.}
	\label{F_wh_sine}
  \end{center}
\end{minipage}\hfill 
\end{figure}

In this section we follow to Ref.\cite{Dzhunushaliev:2009yw}. Let us consider a gravitating system with one phantom scalar field $\varphi$ with the Lagrangian
\begin{equation}
\label{lagrangian}
  L =-\frac{R}{16\pi G}-
      \frac{1}{2}\partial_\mu \varphi \partial^\mu
        \varphi -V(\varphi)~,
\end{equation}
where $R$ is the scalar curvature, $G$ is the Newton's gravitational constant, 
and $V$ is the  Sine-Gordon potential with the reversed sign
\begin{equation}
\label{pot_mex2}
	V=\frac{m^4}{\lambda} \left[
		\cos\left(\frac{\sqrt{\lambda}}{m}\varphi\right)-1
	\right]
\end{equation}
where $m$ is a mass of the Sine-Gordon scalar field, and $\lambda$ is a coupling constant. The corresponding energy-momentum tensor is 
\begin{equation}
\label{emt}
    T^k_i=
    -\partial_i \varphi \partial^k \varphi-
        \delta^k_i \left[
            -\frac{1}{2}\partial_\mu \varphi \partial^\mu
            \varphi-V(\varphi)
        \right]~,
\end{equation}
and corresponding field equations are 
\begin{eqnarray}
\label{Einstein-gen}
    G_{i}^k &=&  8\pi G T^k_i,
\\
\label{field-gen}
	\frac{1}{\sqrt{-g}}\frac{\partial}{\partial 	x^\mu}
	\left(
		\sqrt{-g}\,\, g^{\mu\nu} \frac{\partial \varphi}{\partial x^\nu}
	\right) &=& \frac{\partial V}{\partial
	\varphi}.
\end{eqnarray}
We search the wormhole solution in the form
\begin{equation}
\label{metric_wh}
	ds^2=e^{2 F(r)}dt^2-\frac{dr^2}{A(r)}-(r^2+r_0^2)(d\theta^2+\sin^2\theta d\phi^2).
\end{equation}
Introducing  new dimensionless variables $\phi=(\sqrt{\lambda}/m)\varphi,\, x=m r$ and after some algebraical manipulations one can obtain from \eqref{emt} - \eqref{field-gen} the following equations
\begin{eqnarray}
\label{ein_wh_sine}
	-\frac{A^\prime}{A}x+\frac{1}{A}+\frac{x^2}{x^2+x_0^2}-2&=&
	\frac{x^2+x_0^2}{A}\beta\left(-\frac{A}{2}\phi^{\prime 2}
	+\cos\phi-1\right),
\\
\label{ein_wh2_sine}
	\frac{2 x^2}{x^2+x_0^2}-2-\frac{A'}{A}x+2 x F'&=&
	-\beta(x^2+x_0^2)\phi^{\prime 2},
\\
	\phi^{\prime\prime}+\left(-\frac{\beta}{2}\frac{x^2+x_0^2}{x}\phi^{\prime 2}
	+\frac{A^\prime}{A}+\frac{1}{x}+\frac{x}{x^2+x_0^2}\right)\phi^\prime&=&
	\frac{1}{A}\sin\phi,
\label{field_wh_sine}
\end{eqnarray}
where $\beta=m^2/\lambda$. For the wormhole solution we have to have 
\begin{equation}
\label{bound}
	A'(0) = F'(0) = \phi(0) = 0.
\end{equation}
Substituting \eqref{bound} into \eqref{ein_wh_sine} - \eqref{field_wh_sine} we obtain constraints on the boundary values of metric and scalar filed 
\begin{equation}
\label{ini_wh_sine}
	A(0)=1+2 \beta x_0^2, \quad \phi(0)=\pi, \quad
	\phi^\prime(0)=\sqrt{\frac{2}{\beta x_0^2}}.
\end{equation}
Solving  numerically equations \eqref{ein_wh_sine} - \eqref{field_wh_sine} with the boundary conditions \eqref{ini_wh_sine} and with $\beta=1$, one can obtain the results presented in Fig.\'s  \ref{phi_wh_sine}, \ref{A_wh_sine} and \ref{F_wh_sine}. One can see that asymptotically (at $x\rightarrow \pm \infty$) $\phi \rightarrow \pi$ and $e^{2 F(x)} \rightarrow  \frac{2}{3} x^2$. The last asymptotic says us that asymptotically the wormhole is the anti-de Sitter space. 

\subsection{Ricci flow and topology change}

In this section we would like to show that the Ricci flow starting from the wormhole obtained in subsection \ref{wh} leads to a singularity by $\lambda \rightarrow \lambda_0$. The standard surgery operation applying at the point A (where the singularity appears) leads to the change of topology. 

Let us consider 3D part of the metric \eqref{metric_wh}
\begin{equation}
\label{3-10}
	dl^2 = e^{2u(r, \lambda)} dr^2 + e^{2v(r, \lambda)} 
		\left( d\theta^2+\sin^2\theta d\phi^2
	\right)
\end{equation}
where the initial conditions by $\lambda = 0$ gives us 3D part of the wormhole metric \eqref{metric_wh}
\begin{eqnarray}
	u(r, 0) &=& 0, 
\label{3-20}\\
	e^{2v(r, 0)} &=& A(r).
\label{3-30}
\end{eqnarray} 
According to section \ref{ricci} the Ricci flow is 
\begin{eqnarray}
	\frac{\partial u}{\partial \lambda} &=& 
	2 e^{-2u} \left(
		v'' - u' v' + {v'}^2
	\right), 
\label{3-40}\\
	\frac{\partial v}{\partial \lambda} &=& 
		e^{-2u} \left( v'' - u' v' + 2 {v'}^2 \right) - e^{-2 v}
\label{3-50}
\end{eqnarray}
where $u = u(r, \lambda), v = v(r, \lambda)$. Let us note that there is a soliton solution which is defined as 
\begin{equation}
	\frac{\partial u}{\partial \lambda} = 
	\frac{\partial v}{\partial \lambda} = 0. 
\label{3-60}
\end{equation}
In this case the solution of Eq's \eqref{3-40} \eqref{3-50} is 
\begin{equation}
	\left( e^v \right)' = \pm e^u.
\label{3-70}
\end{equation}
One can show that introducing new coordinate $x$ 
\begin{equation}
	\pm e^u dr = \left( e^v \right)' dr = dx 
\label{3-80}
\end{equation}
the metric \eqref{3-10} can be written in the form 
\begin{equation}
	dl^2 = dx^2 + x^2 (d\theta^2+\sin^2\theta d\phi^2).
\label{3-90}
\end{equation}
Immediately we see that it is 3D Euclidean space. 

\section{Numerical solution}

\begin{figure}[h]
\begin{minipage}[t]{.45\linewidth}
  \begin{center}
  \fbox{
  \includegraphics[height=5cm,width=7cm]{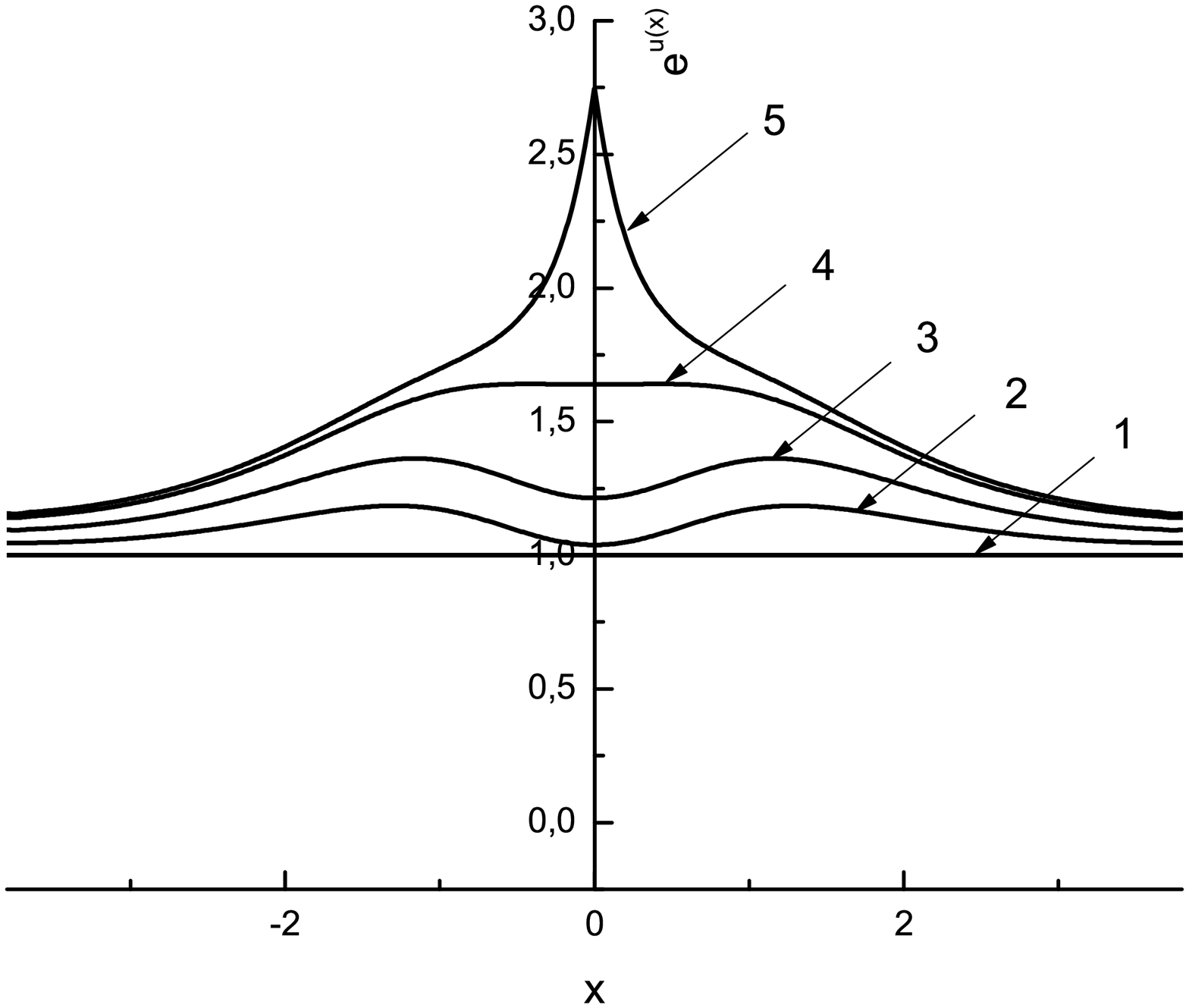}}
  \caption{The profiles of the functions $e^{u(r, \lambda)}$. 
  1 -- by $\lambda = 0$;
  2 -- by $\lambda = 0.3 \cdot \lambda_0$;
  3 -- by $\lambda = 0.6 \cdot \lambda_0$;
  4 -- by $\lambda = 0.9 \cdot \lambda_0$;
  5 -- by $\lambda \approx \lambda_0$.}
  \label{u_x}
  \end{center}
\end{minipage}\hfill
\begin{minipage}[t]{.45\linewidth}
  \begin{center}
  \fbox{
  \includegraphics[height=5cm,width=7cm]{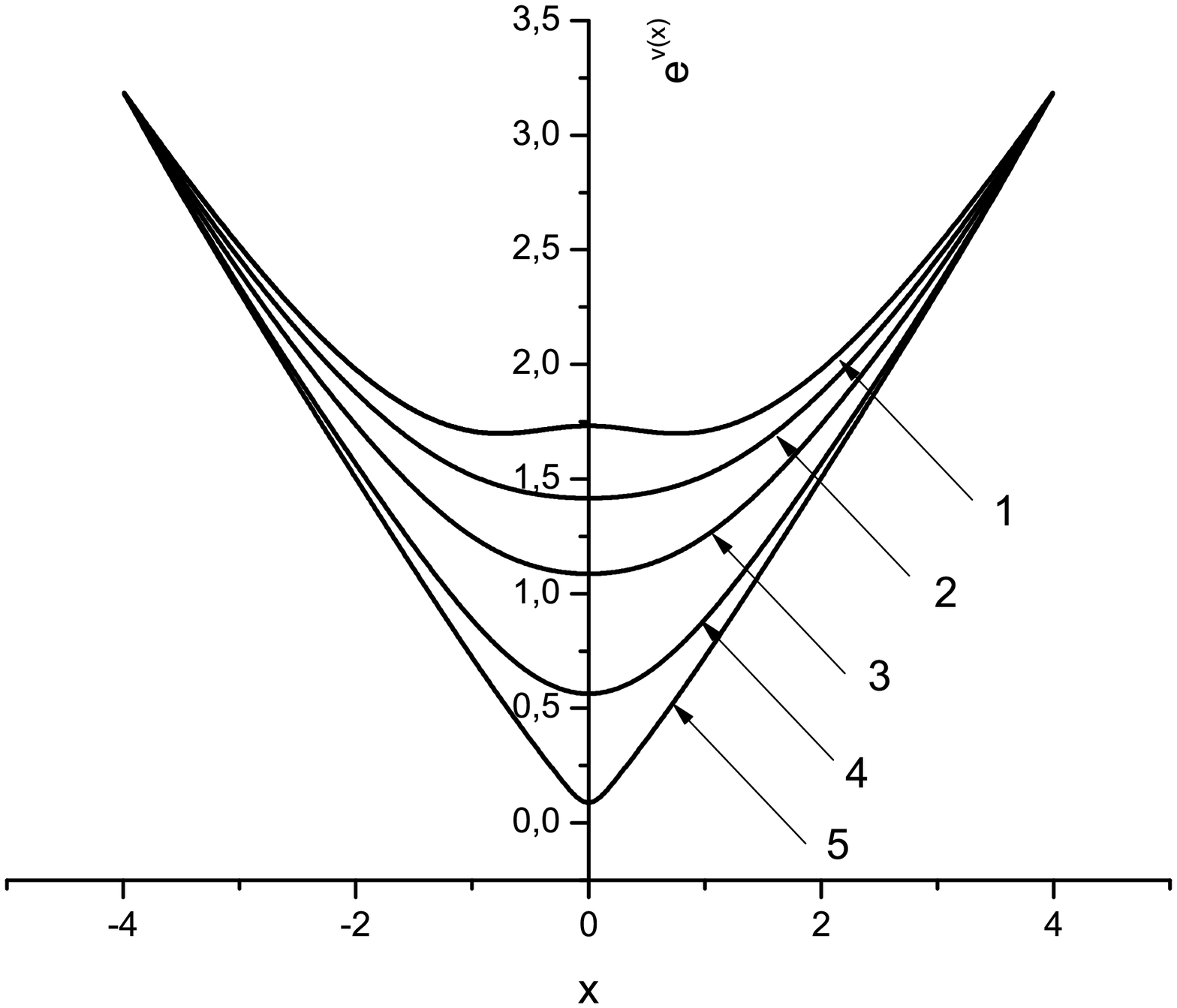}}
	  \caption{The profiles of the functions $e^{v(r, \lambda)}$. 
  1 -- by $\lambda = 0$;
  2 -- by $\lambda = 0.3 \cdot \lambda_0$;
  3 -- by $\lambda = 0.6 \cdot \lambda_0$;
  4 -- by $\lambda = 0.9 \cdot \lambda_0$;
  5 -- by $\lambda \approx \lambda_0$.}
  \label{v_x}
  \end{center}
\end{minipage}\hfill 
\end{figure}

There is a doubt that the analytical solution of Eq's \eqref{3-40} \eqref{3-50} does exist. Therefore we searsch for the analytical solution of Eq's \eqref{3-40} \eqref{3-50}. The initial and boundary conditions in our investigation are 
\begin{eqnarray}
	u(r, 0) &=& 0; \quad 
	u(\infty , \lambda) = 0; 
\label{4-10}\\
	v(r, 0) &=& \frac{1}{2} \ln A(r); \quad 
	\frac{\partial v(0, \lambda)}{\partial r} = 0; \quad 
	\left. v(r , \lambda) \right|_{r \rightarrow \infty} = 
	\frac{1}{2} \left. \ln A(r)\right|_{r \rightarrow \infty}.
\label{4-20}
\end{eqnarray}
Let us note that we consider $\mathcal Z_2$ symmetry solution, i.e. 
$u(-r, \lambda) = u(r, \lambda)$ and $v(-r, \lambda) = v(r, \lambda)$. For the numerical calculations there exist following refinements. Equation \eqref{field_wh_sine} has the term $1/x$ and consequently we have to start the numerical calculations not from $r=0$ but from some $r = \delta \neq 0$. Therefore the numerical calculations for the equations set \eqref{3-40} \eqref{3-50} are made for $\delta \leq |r| \leq x1$. Thus the initial and boundary conditions have the form 
\begin{eqnarray}
	u(r, \delta) &=& 0; \quad 
	u(x1 , \lambda) = 0; 
\label{4-30}\\
	v(r, 0) &=& \frac{1}{2} \ln A(r); \quad 
	\frac{\partial v(\delta, \lambda)}{\partial r} = 0; \quad 
	v(x1 , \lambda) = \frac{1}{2} \ln A(x1).
\label{4-40}
\end{eqnarray}
The result of numerical calculations for the Ricci flow is presented in Fig. \ref{u_x} and \ref{v_x}. From Fig's \ref{v_x} we see that the radius of the wormhole 
$e^{v(0, \lambda)} \stackrel{\lambda \rightarrow \lambda_0}{\longrightarrow} 0$. Simultaneously at this point a singularity is created because 
$e^{u(0, \lambda)} \stackrel{\lambda \rightarrow \lambda_0}{\longrightarrow} \infty$. 

\section{Conclusions}

The question about the topology change in gravity is the challenge for modern theoretical physics. The question is: does exist or not the topology change in gravity. If yes then how one can describe this process. Generally accepted point of view is that the process should be described on the path integral language. Here we investigate the idea that the problem of the topology change can be not connected with the standard quantization scheme of a field theory. It is a independent problem which is connected with the Ricci flows. In this approach the probability of every metric between a static wormhole and final state with the singularity is determined by a positive functional determined by the Ricci flow. 

Here we have shown that starting from a static wormhole created by a phantom Sine-Gordon scalar field we obtain the Ricci flow. The Ricci flow describes a statistical system with microscopical states. Every microscopical state is a metric from the Ricci flow enumerated by a the parameter $\lambda$. The probability of every microscopical state is connected with this parameter. 

\section*{Acknowledgements}

VD am grateful to the Research Group Linkage Programme of the Alexander von Humboldt Foundation for the support of this research.


\begin{thebibliography}{99}

\bibitem{wheel1}
J. Wheeler, 
Ann. of Phys., {\bf 2}, 604(1957); \\
J.A. Wheeler: 
{\sl Geometrodynamics}, Academic Press, USA (1962).

\bibitem{Hamilton} 
R. Hamilton, 
J. Diff. Geom., \textbf{17}, 255 (1982).

\bibitem{Perelman} 
G. Perelman, 
preprint [arxiv:math.DG/0211159]; \\
preprint [arxiv:math.DG/0303109].

\bibitem{geroch0} 
Y. Choquet-Bruhat and R. Geroch, 
Commun. Math. Phys. \textbf{14}, 329-350 (1969).

\bibitem{borde} 
A. Borde,  
Bull. Astr. Soc. India {\bf 25}, 571-577 (1997).

\bibitem{horowitz} 
G.T. Horowitz, 
Class. Quant. Grav. {\bf 8}, 587-601 (1991).

\bibitem{dowkergarcia} 
H.F. Dowker and R.S. Garcia, 
Class. Quant. Grav. {\bf 15}, 1859-1879 (1998); \\
H.F. Dowker, 
{\it Topology change in quantum gravity}, in
\textit{``The future of theoretical physics and cosmology''}, eds. G.W. Gibbons,
S.J. Rankin, E.P.S. Shellard, Cambridge Univ. Press, p.879, (2003).

\bibitem{Sorkin} 
R.D. Sorkin and S. Surya, 
Int.\,J.\,Mod.\,Phys. A{\bf 13}, 3749-3790 (1998). 
  
\bibitem{Csizmadia:2009dm}
  P.~Csizmadia and I.~Racz,
  Class.\ Quant.\ Grav.\  {\bf 27}, 015001 (2010)
  [arXiv:0911.2373 [gr-qc]].

\bibitem{Hirsch}
Morris W. Hirsch, 
''Differential topology``, (Springer-Verlag, New-York-Heidelberg-Berlin, 1976).

\bibitem{topping}
P. Topping, 
{\it Lectures on the Ricci Flow}, 
London Mathematical Society Lecture Notes Series {\bf 325}, 
Cambridge University Press (2006).

\bibitem{Dzhunushaliev:2009yw}
  V.~Dzhunushaliev and V.~Folomeev,
  ``Creation/annihilation of wormholes supported by the Sine-Gordon phantom
  (ghost) field,''
  arXiv:0909.2739 [gr-qc].
  
\bibitem{Dzhunushaliev:2008cz}
  V.~Dzhunushaliev,
  Int. J. Geom. Meth. Mod. Phys., \textbf{6}, 1033-1046 (2009);  
  arXiv:0809.0957 [gr-qc].

\end{thebibliography}
\end{document}